\documentclass[conference]{IEEEtran}
\IEEEoverridecommandlockouts

\usepackage{cite}
\usepackage{amsmath,amssymb,amsfonts}
\usepackage{algorithm}
\usepackage{algorithmic}
\usepackage{graphicx}
\usepackage{textcomp}
\usepackage{xcolor}
\usepackage{float} 
\usepackage{booktabs} 
\usepackage[compatibility=false]{caption} 
\usepackage{bbold}   
\usepackage{subcaption} 
\usepackage{tikz}
\usepackage{lipsum} 
\usepackage{multirow}

\def\BibTeX{{\rm B\kern-.05em{\sc i\kern-.025em b}\kern-.08em
    T\kern-.1667em\lower.7ex\hbox{E}\kern-.125emX}}
\begin{document}

\title{Video QoE Metrics from Encrypted Traffic: Application-agnostic Methodology
\thanks{This research was funded by the ENTM consortium of the Israel innovation authority.}
}

\author{
\IEEEauthorblockN{Tamir Berger, Jonathan Sterenson, Raz Birman, and Ofer Hadar}
\IEEEauthorblockA{School of Electrical and Computer Engineering\\
Ben-Gurion University of the Negev\\
Beer-Sheva, Israel}
}

\maketitle

\begin{abstract}
Instant Messaging-Based Video Call Applications (IMVCAs) and Video Conferencing Applications (VCAs) have become integral to modern communication. Ensuring a high Quality of Experience (QoE) for users in this context is critical for network operators, as network conditions significantly impact user QoE. However, network operators lack access to end-device QoE metrics due to encrypted traffic. Existing solutions estimate QoE metrics from encrypted traffic traversing the network, with the most advanced approaches leveraging machine learning models. Subsequently, the need for ground truth QoE metrics for training and validation poses a challenge, as not all video applications provide these metrics. To address this challenge, we propose an application-agnostic approach for objective QoE estimation from encrypted traffic. Independent of the video application, we obtained key video QoE metrics, enabling broad applicability to various proprietary IMVCAs and VCAs. To validate our solution, we created a diverse dataset from WhatsApp video sessions under various network conditions, comprising 25,680 seconds of traffic data and QoE metrics. Our evaluation shows high performance across the entire dataset, with 85.2\% accuracy for FPS predictions within an error margin of two FPS, and 90.2\% accuracy for PIQE-based quality rating classification.
\end{abstract}

\begin{IEEEkeywords}
QoE, QoE estimation, video conferencing, encrypted traffic, machine learning, WhatsApp.
\end{IEEEkeywords}

\section{Introduction}
Instant messaging-based video call applications (IMVCAs) and Video Conferencing Applications (VCAs) have become essential tools in modern communication transforming how we connect over distances. The COVID-19 pandemic has further accelerated the adoption of video conferencing across various domains, including work, healthcare, education, and personal interactions. IMVCAs have facilitated face-to-face video communication, primarily serving interpersonal communication needs. In addition, Their diverse features have enabled them to penetrate markets traditionally dominated by VCAs. For instance, a study highlighted WhatsApp's utility in education during the COVID-19 pandemic, demonstrating its effectiveness by leveraging various features including video calls, group chats, document sharing, and more\cite{WhatsAppEducation}. As real-time video communication continues to gain popularity, ensuring a high Quality of Experience (QoE) for users across the variety of VCAs and IMVCAs is crucial. Studies \cite{3,7} have shown that network conditions, such as throughput, latency, and packet loss, directly impact user-perceived experience. Traffic patterns can vary significantly, leading to various disruptions across different applications. Therefore, it is critical for network operators to diagnose and address QoE issues as part of efficient network resource management, ensuring a satisfactory user experience. For this purpose, network operators must estimate user QoE using passive measurements of the encrypted traffic traversing their network.
Recent years have seen an expansion in research on video QoE inference from encrypted traffic. Studies\cite{1,2} have focused on estimating QoE from encrypted traffic for streaming services over HTTP. These studies developed methods for QoE monitoring based on machine learning (ML) models that infer Key Performance Indicators (KPIs) from observable traffic. However, there is a fundamental difference between video-on-demand (VoD) and real-time communication. VoD services can buffer content and reduce playback interruptions, whereas real-time communication requires instantaneous data exchange, necessitating different QoE estimation approaches.
Recent studies have focused on developing methods to estimate VCA QoE from encrypted traffic. Sharma et al.\cite{5} explored the estimation of video QoE metrics without using application headers over WebRTC-based VCAs. Similarly, Michel et al.\cite{6} developed methods for passive measurement of Zoom performance in production networks, addressing the challenges posed by proprietary protocols and encrypted headers.
Despite existing research on VCAs, IMVCAs have received relatively little scholarly attention.
In this study, we address the challenges faced by Internet providers in evaluating user QoE using encrypted traffic data. We propose an approach to evaluate video quality metrics from encrypted traffic, building on the existing solutions of Sharma et al.\cite{5} and incorporating additional tools for broader applicability across various platforms. Existing solutions often rely on supervised machine learning models that require samples of quality metrics they aim to predict. This dependency limits their application to platforms that either provide quality metrics or utilize open source communication protocols that allow access to such metrics. However, many applications do not provide QoE metrics or use proprietary protocols, making it challenging to derive quality indicators. To address this limitation, we have developed methods to derive video quality indicators directly from the video itself. By analyzing encrypted traffic and using supervised machine learning models, we estimate quality metrics based on encrypted data. We validated our approach on WhatsApp IMVCA. WhatsApp neither provides quality metrics nor publicly disclose its communication protocol, emphasizing the necessity of our approach. 
Our contributions are as follows:
\begin{itemize}
    \item We develop an application-agnostic framework for estimating video QoE metrics from encrypted traffic.
    \item We conduct a pioneering study on WhatsApp, creating datasets of encrypted traffic and corresponding QoE metrics, thereby closing the gap in IMVCAs.
\end{itemize}
The remainder of this paper is structured as follows: Section II reviews related work in the field of video QoE estimation. Section III details our proposed methodology. Section IV presents the experimental setup and evaluation results. Section V discusses the implications of our findings, and Section VI concludes the paper with future research directions.

\section{Methodology}
Our primary objective is to utilize features derived from encrypted packet traffic streams to predict QoE metrics using ML-based models. Our work extends the approach of Taveesh Sharma et al.\cite{5}, making it applicable to all proprietary UDP-based video communication applications. This is accomplished by providing alternative QoE metrics for which the inference of ground truths, including data collection for learning or validation, is always feasible.
As an ML-driven approach, it requires the collection of datasets containing both traffic traces—the inputs—and the corresponding ground truth QoE metrics labels for each time slot. We proceed to provide details of our QoE metrics and the generation and characterization of the dataset. Then, we describe the ML models employed for QoE prediction.

\subsection{Dataset collection method}\label{AA}
To obtain a dataset that includes our specifically selected QoE indicators and ensures accurate time synchronization with the encrypted packet traffic data, we developed an appropriate data collection method. The current implementation supports conversations between two participants. To execute the method, two distinct roles must be performed by different parties in the conversation: receiver and transmitter. Each participant can simultaneously function as both transmitter and receiver. Data collection is executed on the receiver's side and involves several parallel components:
1. Capturing all traffic data during the call. At the end of each session, specific data, including time, IP, UDP, and RTP headers, is extracted.
2. Capturing two types of screen captures: The first type consists of full-screen recordings at a low rate of one frame per second, captured in high-quality, lossless format. The second type involves high-speed screen captures of a small, focused window of the screen without quality demands. Both types of screen recordings are essential for inferring QoE metrics, which will be elaborated in the following sections.

\subsection{WhatsApp Dataset}
Utilizing the data collection method, we gathered data from 107 WhatsApp video sessions, totaling 25,680 seconds, with each session lasting 4 minutes. To develop a general model, the video sessions were conducted under various network conditions: (1) \textit{Packet Loss}, Sessions experienced packet loss in varying percentages. (2) \textit{Bandwidth Limits}, Fixed bandwidth limits were imposed throughout the sessions. (3) \textit{Bandwidth Drops}, unexpected bandwidth drops were introduced during the sessions. For sessions with bandwidth limits, the bandwidth was set to one of the following values: 250 kBps, 125 kBps, 60 kBps, 30 kBps, or 15 kBps, evenly distributed among the sessions. At 15 kBps, although the conversation quality is very low, intelligible communication is still available. Sessions with bandwidth drops experienced sudden decreases in bandwidth, from approximately 250 kBps to a variable bandwidth in range 10 to 150 kBps, causing various communication interruptions, primarily frame freezes and subsequent reduced quality. For sessions with packet loss, the calls experienced varying loss percentages of 0\%, 1\%, 2\%, 5\%, and 10\%, uniformly distributed, with a fixed bandwidth of 250 kBps. The video sessions were conducted with participants' faces visible on both sides of the conversation, simulating typical face-to-face video calls.

\subsection{Video stream classification}
\begin{figure}[htb]
\centering
\includegraphics[width=\linewidth]{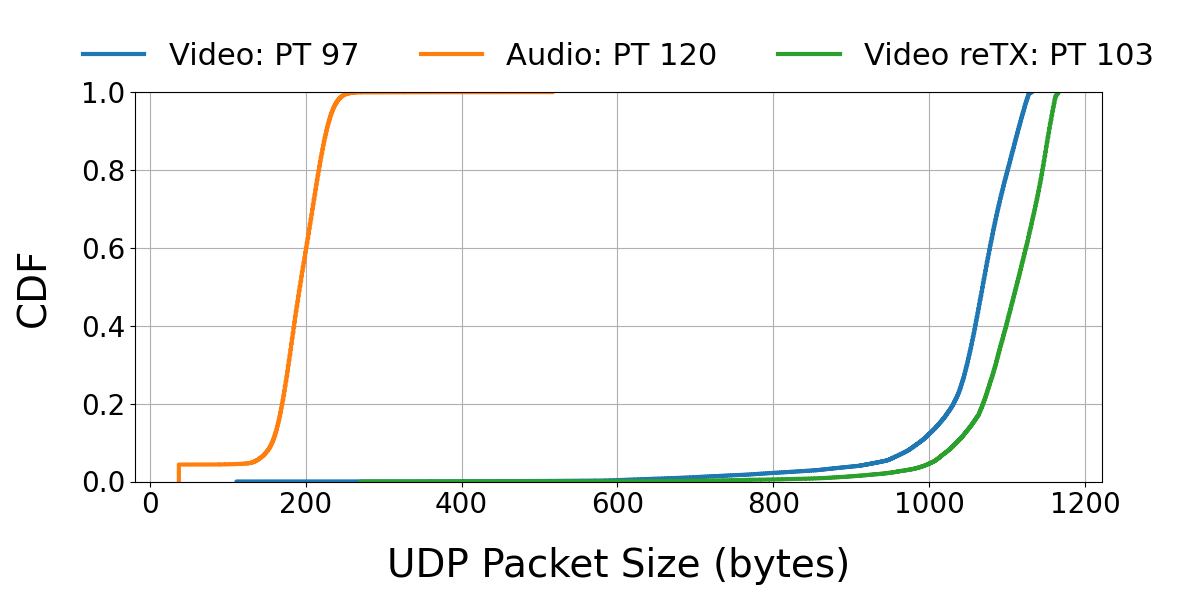}
\caption{Cumulative distribution function (CDF) of UDP packet size by payload type for the WhatsApp dataset. PT denotes payload type and reTX denotes retransmission.}
\label{fig:udp_size_cdf_plot}
\end{figure}

The classification process applied to the collected data consisted of two steps. First, detect the packets related to the specific application stream. Second, isolate the video stream data from these packets. We identify the video application traffic by the packets' IP addresses and subsequently extract only the video packet data, disregarding all other packet types. Previous studies on Zoom\cite{6, 13} have distinguished media types by utilizing RTP headers, specifically the RTP payload type, to classify packets into different categories, such as video, screen sharing, and audio.
In study \cite{5}, Taveesh Sharma et al. differentiated packets by size, based on the assumption that audio packets require fewer bits than video packets. Adopting this approach, we classify packets based on their size, ensuring the method's applicability even when RTP headers are encrypted, or the payload type has been modified by application protocol changes.
Upon investigating our WhatsApp dataset, we found that it uses RTP payload type (PT) 97 for video, PT 120 for audio, and PT 103 for video retransmission throughout the entire dataset. The ground truth distribution of packet lengths by RTP payload type in our entire dataset is shown in Figure~\ref{fig:udp_size_cdf_plot}.
Our analysis revealed that the most effective classification threshold is 275 bytes, achieving a classification accuracy of 99.99\%. Table \ref{tab:tab1} presents the confusion matrix for video packet classification.

\begin{table*}[htbp]
\caption{Classification Accuracy of WhatsApp Packets}
\centering
\begin{tabular}{|c|c|c|c|c|}
\hline
\textbf{Actual}& \multicolumn{2}{|c|}{\textbf{Classified}} & \textbf{Packet Count} & \textbf{Average Packet Size (bytes)} \\
\cline{2-3}
\textbf{} & \textbf{Non-Video} & \textbf{Video} & & \\
\hline
\textbf{Non-video} & 99.93\% & 0.07\% & 285751 & 187 \\
\hline
\textbf{Video} & 0.01\% & 99.99\% & 3583870 & 1054 \\
\hline
\end{tabular}
\label{tab:tab1}
\end{table*}

\subsection{Ground Truth Metrics}
\subsubsection{Frame Rate}
The procedure for determining the frame rate involves the following steps:
\begin{itemize}
    \item \textit{Creating a Unique Visual Identifier for Each Frame on the Transmitter Side:} A screen displaying \( S \) changing signs is placed in the background of the participant. The signs change at rate \( V\), which must be higher than the maximum frame rate of the video application.  The number of signs \( S \) must satisfy  \( S \geq V \cdot T \), where \( T \) is the duration of the time slot in seconds. This ensures that each sign appears at most once within a time slot, making it a unique identifier for each frame.
    \item \textit{Frame Rate Measurement on the Receiver Side:} On the receiver's end, successive window captures are taken of the specific area where the changing sign appears. These captures are performed at a rate \( V \) and are collected along with their corresponding timestamps. Consequently, there are more captures than frames within the time slot, with some duplications, ensuring that each frame is captured at least once.
    \item \textit{Associating Frames with Time Slots Based on Arrival Times:} At the end of each session, the program processes the collected images in chronological order, retaining only the first capture of each frame by removing all duplicates. Thus, each capture in this list represents a unique frame in the session, ensuring that all frames are represented. The captures are then grouped into time slots.
    \item \textit{Calculating Frame Rate for Each Time Slot:} For each time slot, the number of unique captures associated with it is summed. The frame per second rate for time slot \( i \) is calculated as:
    \begin{equation}
        \text{fps}_i = \frac{1}{T} \sum_{k \in i} \mathbb{1}_{[\text{capture}_k \neq \text{capture}_{k-1}]}
    \end{equation}
    where \( \mathbb{1} \) is the indicator function for a new frame, \( T \) is the duration of the time slot in seconds, \( i \) is the time slot number, and \( k \) is the index of the capture.
\end{itemize}
The procedure is described in Algorithm~\ref{alg:frame_rate_inference}.
\begin{algorithm}[b]
\caption{Frame Rate Calculation}
\label{alg:frame_rate_inference}
\begin{algorithmic}
\STATE \textbf{Input:} captures, endTimeSlots
\STATE \textbf{Output:} frameRates
\STATE $N \leftarrow $ captures.length
\STATE uniqueCaptures $\leftarrow [\ ]$ \hspace{1mm} \textit{(empty list)}
\STATE frameRates $\leftarrow \{\}$ \hspace{6mm} \textit{(empty dictionary)}
\FOR {endTime in endTimeSlots}
    \STATE frameRates[endTime] $\leftarrow 0$
\ENDFOR
\FOR {i in 1:N}
    \IF {captures[i].image $\neq$ captures[i-1].image}
        \STATE uniqueCaptures.append(captures[i])
    \ENDIF
\ENDFOR
\FOR {capture in uniqueCaptures}
    \STATE frameRates[$\lceil$ capture.timeStamp $\rceil$] += 1
\ENDFOR
\STATE \textbf{return} frameRates
\end{algorithmic}
\end{algorithm}
To validate the proposed method, we generated 30 distinct videos, each with a constant frame rate ranging from 1 to 30 frames per second (FPS). By concurrently playing these videos and executing the frame rate detection program on the same device, we eliminated network effects, ensuring precise frame rate display on the screen. The validation results demonstrated high accuracy across the entire tested FPS range, with a Mean Absolute Error (MAE) of 0.06 FPS and an accuracy of 97.5\% within an error tolerance of 5\% relative to the ground truth.
The procedure requirement to introduce visual changes in the participant's background ensures high accuracy in frame rate detection but adds complexity for the user during the video session. Practically, this requirement can be removed by leveraging the inherent dynamics within the video content, if present. We are currently developing such an approach. 

\subsubsection{Spatial Quality Assessment}
In order to obtain ground truth spatial quality metrics, we capture lossless screen frames at 1-second rate during the session. For each frame, we calculate quality scores using  two no-reference image quality
assessment models: BRISQUE\cite{brisque} and PIQE\cite{piqe}. Both models provide quality scores ranging from 0 to 100, with lower values indicating fewer distortions. BRISQUE evaluates spatial quality based on natural scene statistics, while PIQE assesses perceptual aspects of local distortions in the spatial domain. For each model, the average score per time slot is used to create a quality score label.
Table \ref{tab:quality_scales} shows the rating scale of PIQE and BRISQUE metrics used to assess the spatial quality rating for each time slot. The ratings are categorized as excellent, good, fair, poor, and bad. For PIQE, the quality scale and respective score ranges are assigned based on the experimental analysis in \cite{piqe_scale}. For BRISQUE, due to the lack of a universally agreed-upon quality scale rating, we propose a linear quality scale.

\begin{table}[t]
\caption{Quality Scale Ratings of PIQE and BRISQUE}
\centering
\begin{tabular}{|c|c|c|}
\hline
\textbf{PIQE Range} & \textbf{BRISQUE Range} & \textbf{Rating} \\
\hline
0 - 20 & 0 - 20 & Excellent \\
\hline
21 - 35 & 21 - 40 & Good \\
\hline
36 - 50 & 41 - 60 & Fair \\
\hline
51 - 80 & 61 - 80 & Poor \\
\hline
81 - 100 & 81 - 100 & Bad \\
\hline
\end{tabular}
\label{tab:quality_scales}
\end{table}

\subsection{Data Analysis}
\begin{figure}[b]
    \centering
    \begin{subfigure}[b]{0.24\textwidth}
        \includegraphics[width=\textwidth]{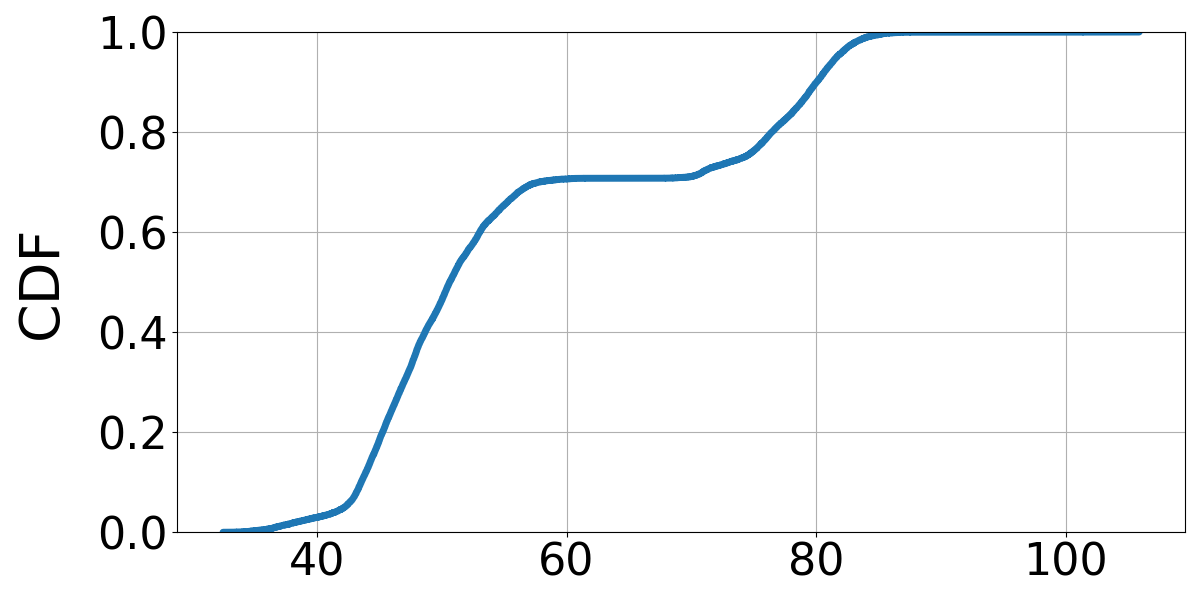}
        \caption{BRISQUE score.}
        \label{fig:cdf_brisque}
    \end{subfigure}
    \begin{subfigure}[b]{0.24\textwidth}
        \includegraphics[width=\textwidth]{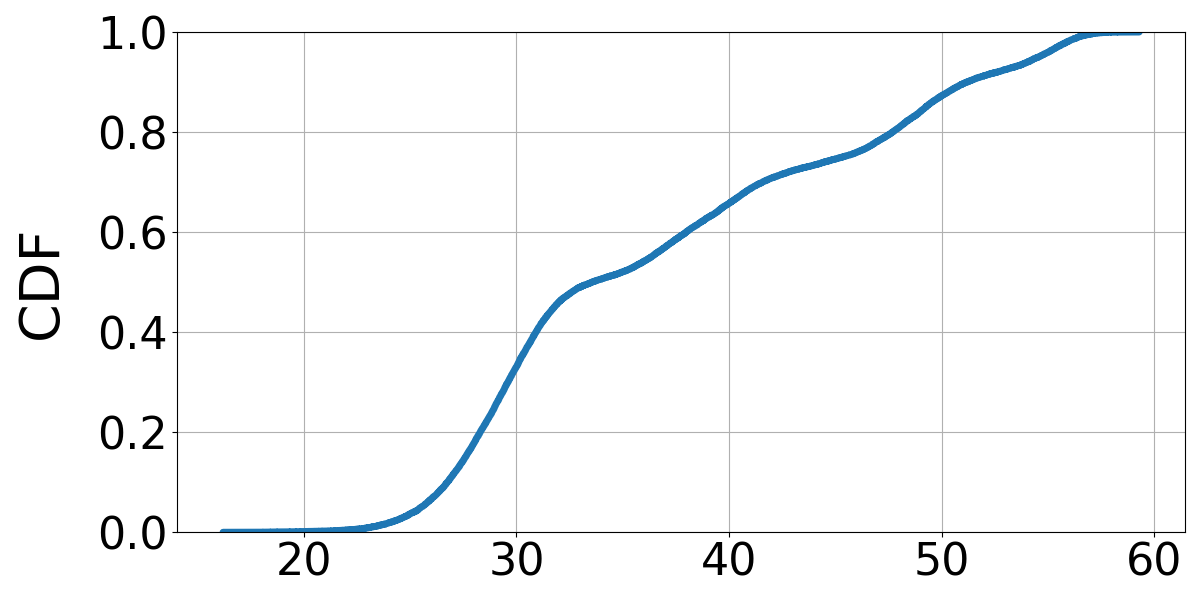}
        \caption{PIQE score.}
        \label{fig:cdf_piqe}
    \end{subfigure}
    \begin{subfigure}[b]{0.24\textwidth}
        \includegraphics[width=\textwidth]{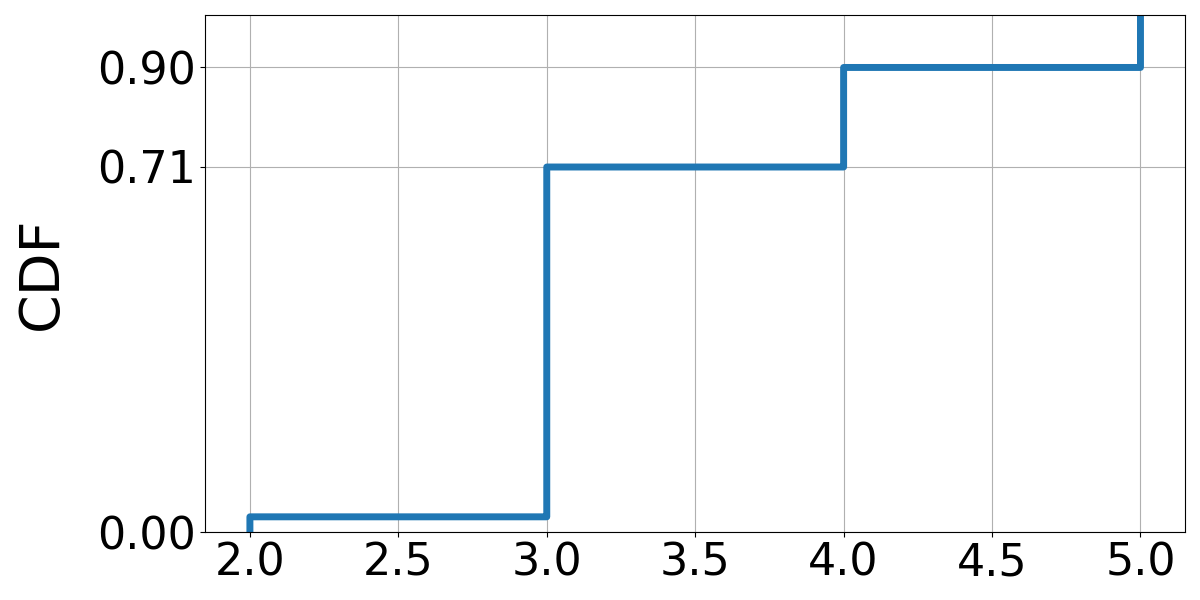}
        \caption{BRISQUE-based rating.}
        \label{fig:cdf_brisque_ratings}
    \end{subfigure}
    \begin{subfigure}[b]{0.24\textwidth}
        \includegraphics[width=\textwidth]{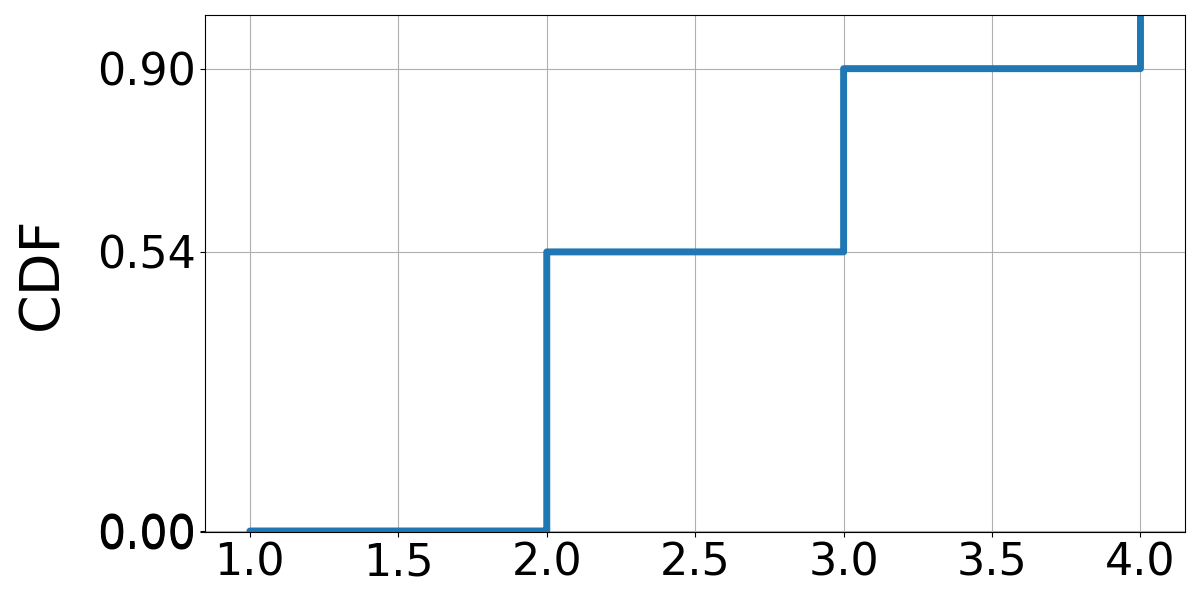}
        \caption{PIQE-based rating.}
        \label{fig:cdf_piqe_ratings}
    \end{subfigure}
    \caption{Characterization of the spatial quality through CDFs.}
    \label{fig:cdf_plots}
\end{figure}

Next, we present an analysis of the collected dataset. Figure~\ref{fig:cdf_plots} illustrates the spatial quality characteristics through cumulative distribution functions (CDF) for BRISQUE score, PIQE score and BRISQUE- and PIQE-based quality ratings.

\begin{figure*}[htb]
    \centering
    \begin{subfigure}[b]{0.32\textwidth}
        \includegraphics[width=\linewidth]{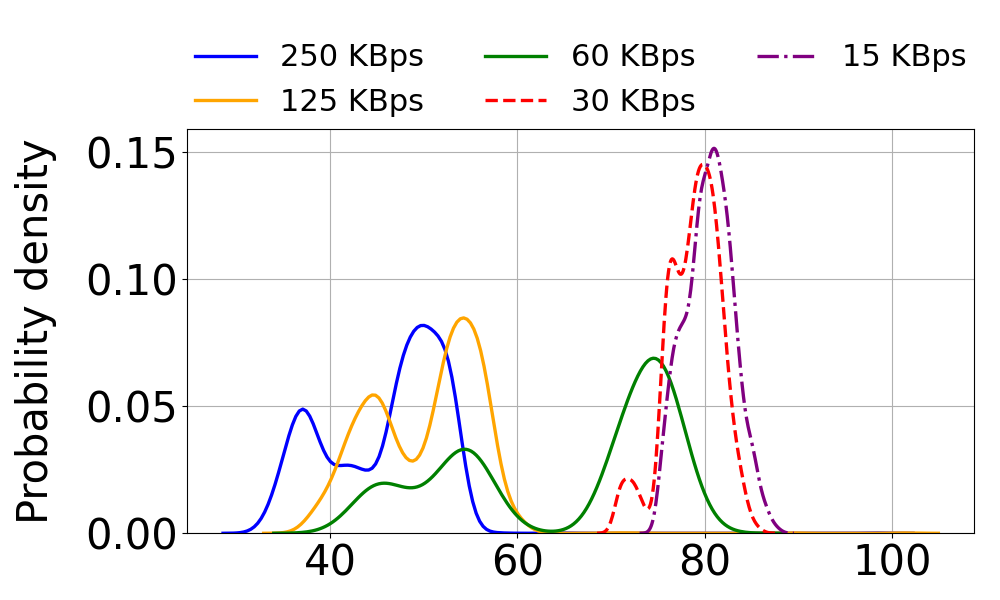}
        \caption{BRISQUE.}
        \label{fig:brisque-bandwidth density}
    \end{subfigure}
    \begin{subfigure}[b]{0.32\textwidth}
        \includegraphics[width=\linewidth]{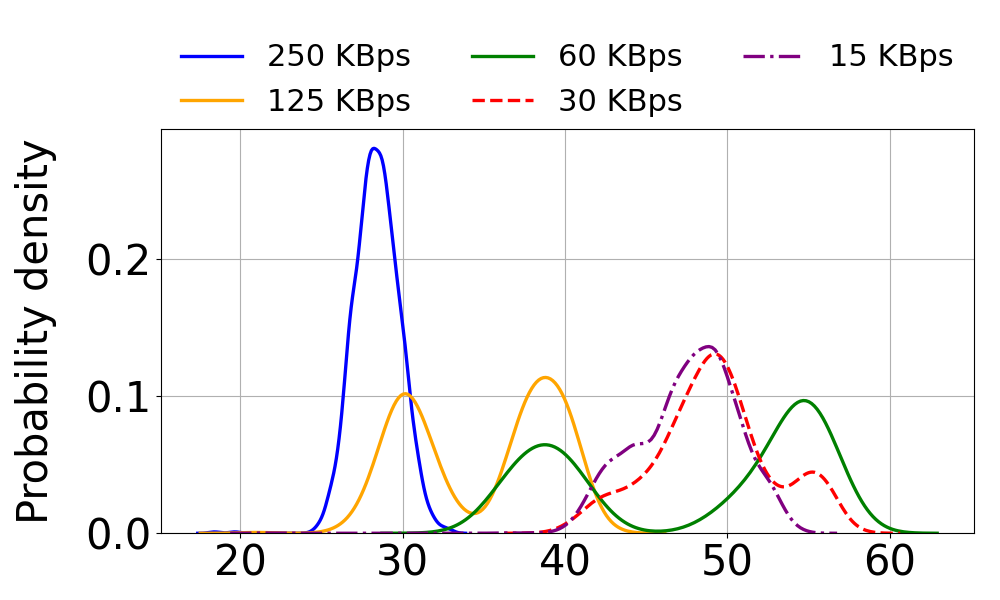}
        \caption{PIQE.}
        \label{fig:piqe-bandwidth density}
    \end{subfigure}
    \begin{subfigure}[b]{0.32\textwidth}
        \includegraphics[width=\linewidth]{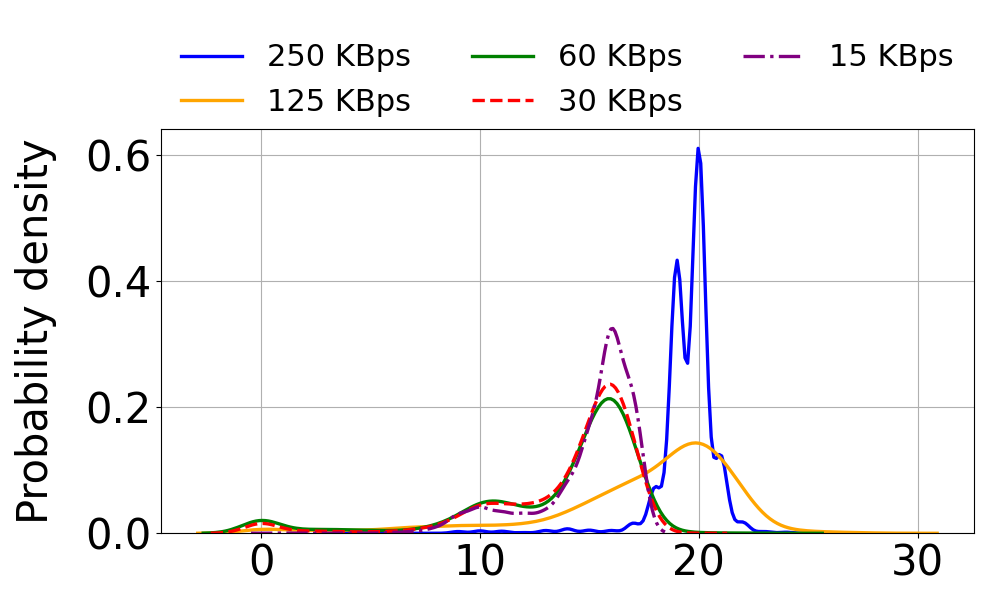}
        \caption{FPS.}
        \label{fig:fps-bandwidth density}
    \end{subfigure}
    \caption{Density distribution of BRISQUE, PIQE and frame rate corresponding to each bandwidth limit: 250 kBps, 125 kBps, 60 kBps, 30 kBps, and 15 kBps.}
    \label{fig:density_by_bandwidth_plots}
\end{figure*}

Figure~\ref{fig:cdf_brisque} depicts the distribution of BRISQUE scores, showing two separate increases that indicate a distinction between two general quality groups obtained during the sessions. Correspondingly, the BRISQUE-based quality ratings are predominantly 'average' (66\%), aligning with the primary increase, while 'poor' and 'bad' ratings together account for 29\%. The 'good' quality rating is rare, comprising only 3\%. The PIQE distribution in figure~\ref{fig:cdf_piqe} shows a more continuous increase, indicating a more even distribution. For the PIQE ratings, most samples are classified as 'Good' (52\%) and 'Fair' (37\%), while the remaining samples are rated as 'Poor' (11\%). As the PIQE-based quality rating is derived from experimental analysis, it offers a more reliable metric compared to the BRISQUE-based rating. 
Figure~\ref{fig:density_by_bandwidth_plots} presents the probability density distributions of BRISQUE, PIQE, and frame rate under various bandwidth limitations, illustrating the strong correlation between spatial image quality and bandwidth in WhatsApp. Figure~\ref{fig:piqe-bandwidth density} shows that the PIQE distribution for each bandwidth limit is concentrated around one or two central values. At 250 kBps, typically used by WhatsApp without network restrictions, the distribution is narrowly centered around a low value, indicating stable high quality. In contrast, at 125 kBps and 60 kBps, the distributions are centered around two distinct values for each bandwidth. This suggests different spatial adaptations employed by WhatsApp in response to bandwidth limitations. Figure~\ref{fig:fps-bandwidth density} shows the frame rate distribution has two primary peaks: the 250 kBps distribution is narrowly centered around 20 FPS, the 125 kBps distribution is more widely centered around 20 FPS, and the 60 kBps, 30 kBps, and 15 kBps distributions are centered around approximately 15 FPS. This finding is consistent with our analysis of WhatsApp logs, which revealed that the intended frame rates for video sessions are primarily 15 or 20 FPS, with actual frame rates fluctuating around these target values. This analysis suggests that WhatsApp prioritizes maintaining stable frame rates of 15 to 20 FPS, while adjusting spatial quality to accommodate varying bandwidth conditions. Under constant bandwidth conditions, WhatsApp efficiently maintains continuous video, with only a minor density below 10 FPS, indicating minimal occurrences of frame freezes.

\begin{figure}[b]
    \centering
    \begin{subfigure}[b]{0.24\textwidth}
        \includegraphics[width=\textwidth]{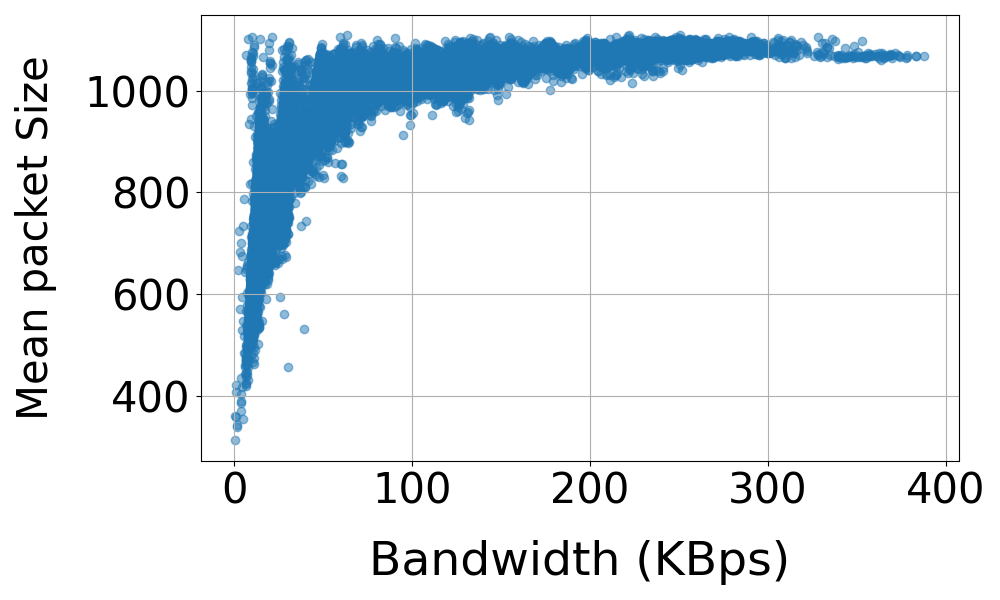}
        \caption{Mean packet size.}
        \label{fig:packet size mean vs bandwidth}
    \end{subfigure}
    \begin{subfigure}[b]{0.24\textwidth}
        \includegraphics[width=\textwidth]{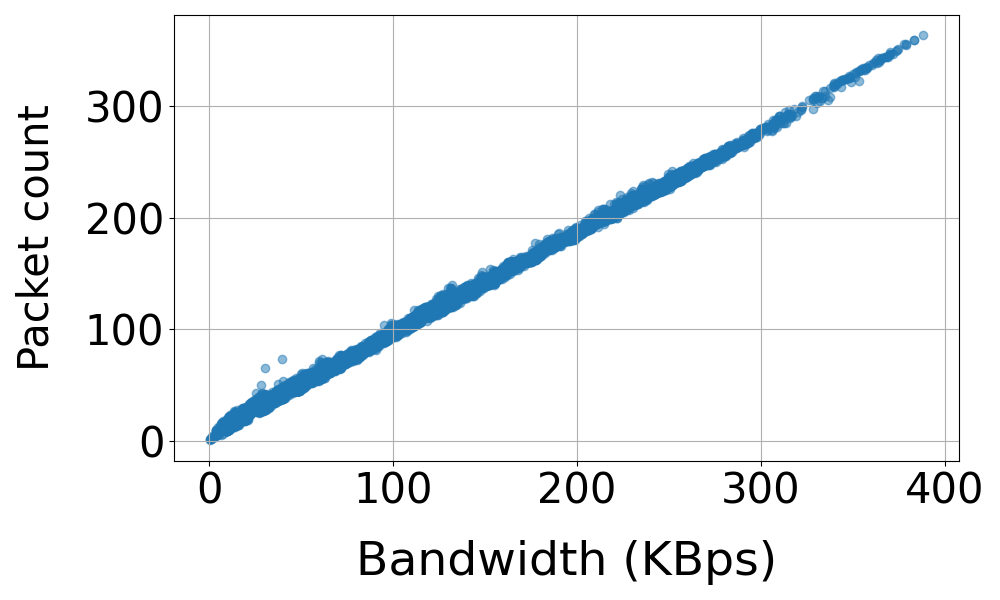}
        \caption{Packet count.}
        \label{fig:packet count vs bandwidth}
    \end{subfigure}
    \caption{Relationship between bandwidth and (a) mean packet size, and (b) packet count.}
    \label{fig:packet vs bandwidth}
\end{figure}

Figure~\ref{fig:packet vs bandwidth} presents the relationship between packet size and packet count with utilized bandwidth. Figure~\ref{fig:packet size mean vs bandwidth} demonstrates a logarithmic relationship between packet size and utilized bandwidth. The majority of packets (82\%) range between 900 to 1150 bytes, indicating consistent packet sizes for most of the utilized bandwidth range. Packet sizes below 700 bytes are infrequent (less than 5\%) and are associated with low utilized bandwidth (below 40 kBps). Figure~\ref{fig:packet count vs bandwidth} demonstrates a linear relationship between packet count and utilized bandwidth, exhibiting perfect correlation.

\subsection{Traffic Features and ML Models}
To extract traffic features, we adopt the approach suggested by Taveesh Sharma et al.\cite{5}, which benefits from lightweight features and enables QoE predictions in scenarios where RTP headers are encrypted. Accordingly, we implemented two models: the UDP model, which utilizes only IP and UDP headers, and the RTP model, which leverages RTP headers in addition to the IP and UDP headers. Packets are grouped according to their arrival times, and traffic features are calculated for each time slot (sample). For the UDP model, we derive 18 features from the IP/UDP headers, based on packet sizes and inter-arrival times. For the RTP model, we derive 11 features from the following RTP headers: timestamps, marker bit, and sequence number. Further details on these features are extensively discussed in \cite{5}.
Previous Research \cite{1, 5, 11, 12} leveraging ML techniques for network data analysis have shown that decision tree-based models, specifically Random Forest, excel as the underlying ML approach. Consequently, we consider both bagging and boosting ensembles based on trees: Random Forest and the gradient boosting model XGBoost.
Model parameters were optimized using standard grid search techniques.
To enhance training efficiency, reduce model complexity, and minimize the risk of overfitting, we prioritized using a small number of trees during parameter tuning, specifically 10 to 100 trees. All evaluations throughout the study are done through 5-fold cross-validation. 
We ensure that the five folds maintain the same ratio between the different network conditions and the various limitations within each condition. 
Following evaluation, Random Forest was selected as it yielded higher accuracy and faster training times.
Our method utilizes two parameters: $T$, which is adjustable, and $V$, which is specific to the application. The parameter $S$ is determined according to $T$ and $V$. In our work, we set $T$ to 1 second and $V$ to 60 FPS, which is twice the maximum frame rate of WhatsApp. Analysis of WhatsApp logs revealed that the default maximum frame rate is 30 FPS. In practice, our frame rate detection method indicates that the majority of the dataset samples (97.4\%) have a frame rate of up to 22 FPS.

\section{Performance evaluation}
This section presents the performance evaluation results of UDP and RTP models on our WhatsApp dataset, focusing on the specified QoE metrics. Table ~\ref{tab:performance_evaluation_mae} presents the MAE values for FPS, BRISQUE, and PIQE metric predictions for both UDP and RTP models. The results indicate an inverse impact of network conditions on spatial and temporal quality. Frame rate estimation shows minimal error under bandwidth limitations, while BRISQUE and PIQE estimations are more accurate under packet loss condition. We further analyze network conditions contributing to model inaccuracies, identify potential error sources, and examine WhatsApp's QoE under varying network conditions.

\begin{table}[b]
\caption{Mean Absolute Error (MAE) of RTP and UDP Models for FPS, BRISQUE, and PIQE Metrics Under Various Network Conditions}
\centering
\begin{tabular}{|c|c|c|c|c|c|c|}
\hline
& \multicolumn{2}{c|}{\textbf{FPS}} & \multicolumn{2}{c|}{\textbf{Brisque}} & \multicolumn{2}{c|}{\textbf{PIQE}} \\
\cline{2-7}
& \textbf{RTP} & \textbf{UDP} & \textbf{RTP} & \textbf{UDP} & \textbf{RTP} & \textbf{UDP} \\
\hline
\textbf{Bandwidth Limits} & 1.27 & 1.46 & 4.72 & 4.36 & 2.42 & 2.36 \\
\hline
\textbf{Bandwidth Drops} & 1.06 & 1.36 & 3.29 & 3.28 & 2.05 & 1.99 \\
\hline
\textbf{Packet Loss} & 3.43 & 3.38 & 3.10 & 2.63 & 1.70 & 1.80 \\
\hline
\end{tabular}
\label{tab:performance_evaluation_mae}
\end{table}

\subsection{Bandwidth Drop Effects}
\begin{figure}[t]
\centering
\includegraphics[width=\linewidth]{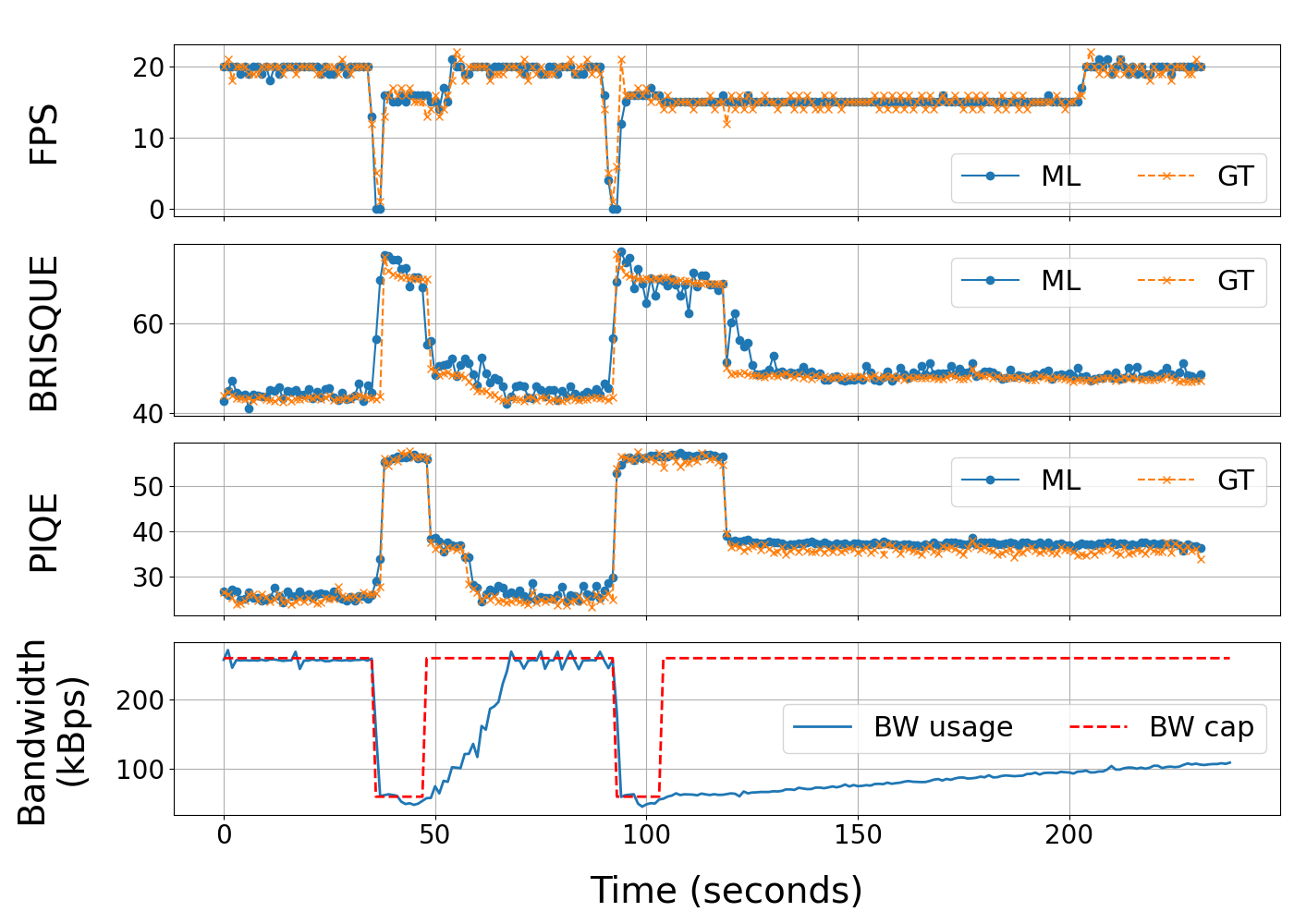}
\caption{QoE metrics predictions during a video session under unexpected bandwidth reductions. From top to bottom, the graphs represent FPS, BRISQUE, PIQE, and bandwidth usage and cap. ML denotes model predictions, GT denotes ground truth and cap denotes limit.}
\label{fig:bandwidth_drops_session}
\end{figure}

Bandwidth drops can significantly degrade QoE, by causing temporal and spatial video interruptions. Next, we analyze the models’s performance under unexpected bandwidth reductions, and examine their impact on the specified video quality metrics.
Figure~\ref{fig:bandwidth_drops_session} presents the UDP model predictions compared to the ground truth values, alongside the measured bandwidth usage and limit during a WhatsApp video session experiencing unexpected bandwidth reductions. It reveals the differences in recovery times between the initial and subsequent bandwidth drops within a single session. Two major bandwidth drops are shown as sharp declines in the measured bandwidth curve, each lasting approximately 10 seconds. Following the second drop, bandwidth utilization exhibited only a moderate and slower increase compared to the rapid recovery observed after the first drop. This behavior demonstrates the recovery adaptations implemented by the WhatsApp protocol. Since QoE is severely impacted by bandwidth drops, the application attempts to maintain a stable bandwidth to sustain a reasonable QoE \cite{17,18,19}.

\subsection{Frame Rate Estimation}
\begin{figure*}[ht]
    \centering
    \begin{subfigure}[b]{0.4\textwidth}
        \includegraphics[width=\linewidth]{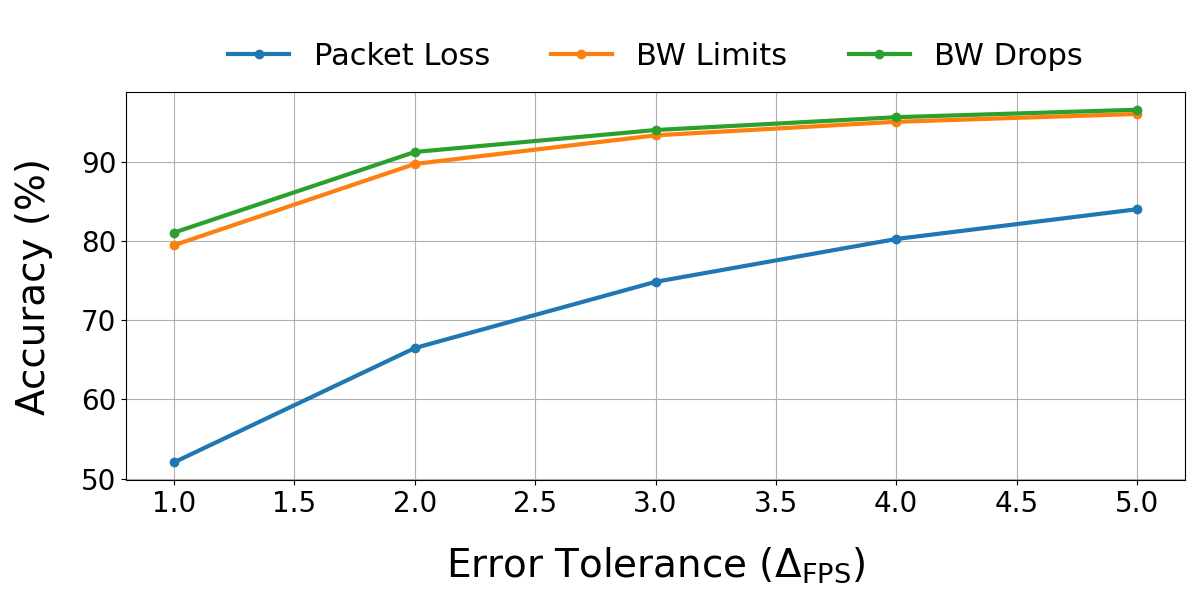}
        \caption{}
        \label{fig:error_tol_fps_all_delta}
    \end{subfigure}
    \begin{subfigure}[b]{0.4\textwidth}
        \includegraphics[width=\linewidth]{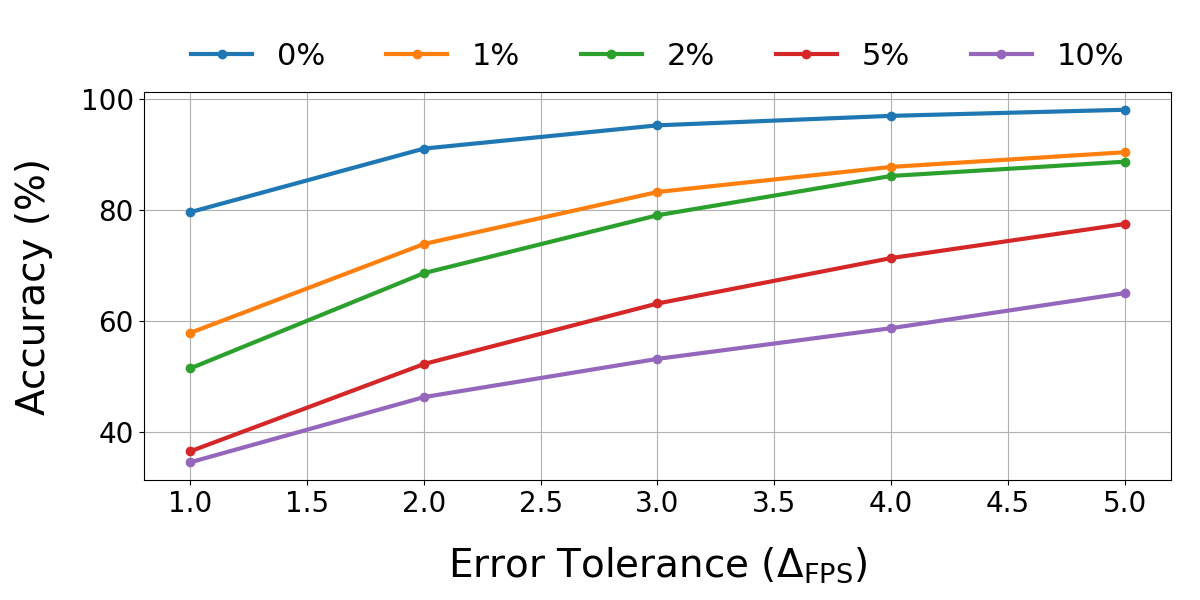}
        \caption{}
        \label{fig:fps_acc_loss}
    \end{subfigure}
    \caption{FPS prediction accuracy versus error tolerance under different network conditions. (a) Comparison of bandwidth limits, bandwidth drops, and packet loss. (b) Impact of various packet loss rates: 0\%, 1\%, 2\%, 5\%, and 10\%.}
    \label{fig:fps_acc}
\end{figure*}

We now examine the prediction performance of frame rate, a crucial metric that directly influences video smoothness and overall QoE. Figure \ref{fig:fps_acc} illustrates the accuracy of frame rate predictions as a function of error tolerance under the described network conditions. Under stable bandwidth limits and bandwidth drop conditions, the frame rate prediction accuracy within a 1 FPS error tolerance is approximately 80\%. This accuracy increases to over 94\% when the error tolerance is extended to 3 FPS. However, under packet loss conditions, the prediction accuracy for a 1 FPS error tolerance is 52\%, reaching 80\% only at a 4 FPS error tolerance. Figure \ref{fig:fps_acc_loss} compares frame rate predictions under different packet loss percentages, highlighting the impact of packet loss on prediction accuracy. As packet loss rates increase, prediction accuracy decreases significantly.

\subsection{Packet Loss Effect}
\begin{figure}[htb]
\centering
\includegraphics[width=\linewidth]{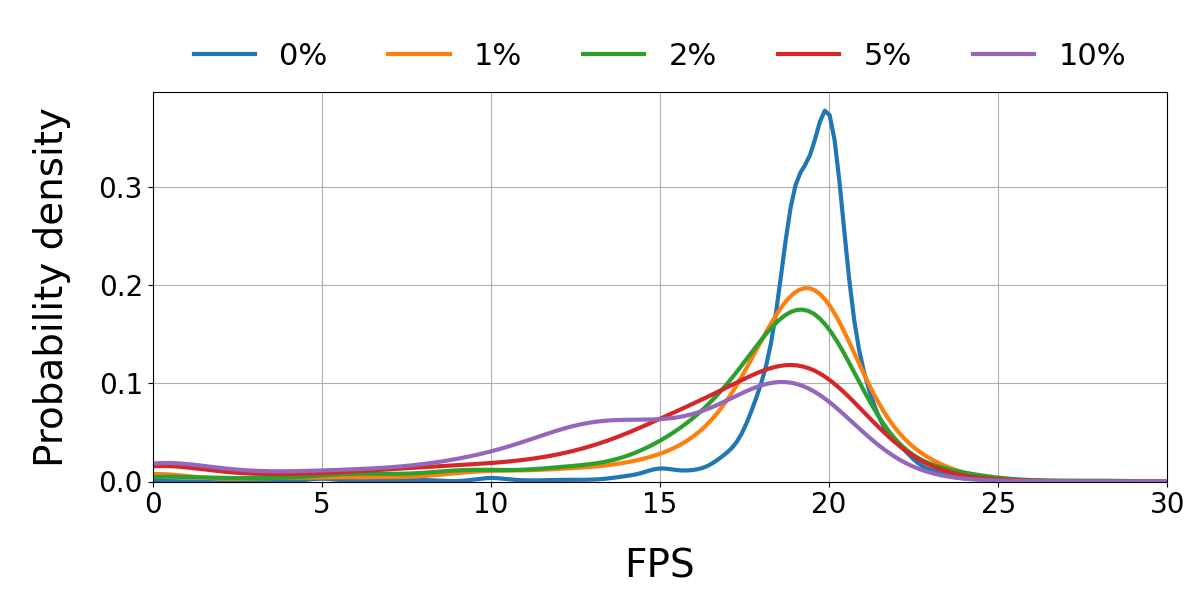}
\caption{Density distribution of frame rate for various packet loss rates: 0\%, 1\%, 2\%, 5\%, and 10\%.}
\label{fig:fps_density_loss}
\end{figure}

The challenge in predicting frame rate during significant packet loss arises from the arrival of incomplete data for frame reconstruction, leading to frame freezes and drops. Previous research indicates that even minor packet loss can affect consecutive frames, directly impacting prediction accuracy, which relies on network traffic features within the same time slot. The fluctuations and inconsistencies in frame delivery under these conditions make it difficult for models to maintain accurate predictions \cite{1,14,21}.
Figure \ref{fig:fps_density_loss} presents the density distribution of frame rates across different packet loss percentages. It demonstrates that as packet loss increases, the density peaks of FPS values decrease and the spread of FPS values becomes wider. The variability and instability in FPS values are particularly evident at higher packet loss rates (5\%, 10\%), further complicating the prediction task.

\subsection{Image Quality estimation}
Next, we present the performance evaluation of the spatial quality metrics BRISQUE and PIQE. We address quality score estimation as a regression task and quality rating estimation as a classification task.
Table~\ref{tab:quality_scale_rating_accuracy} presents the classification accuracy of PIQE and BRISQUE ratings for each described network condition. Both models demonstrated noticeably high performance under packet loss conditions, achieving approximately 99\% accuracy. Our analysis indicates that despite the adverse effects on frame rate, image quality remains largely unaffected under packet loss condition. Under this condition, all samples are rated as 'fair' for the BRISQUE rating, and 98\% of the samples are rated as 'good' for the PIQE rating, thereby simplifying the classification task. These ratings represent the highest quality classes observed for each scale in the entire dataset, indicating WhatsApp's spatial quality stability under packet loss conditions.
We continue to focus on evaluating the PIQE ratings, as they offer two important advantages over the BRISQUE ratings. First, the dataset distribution according to the PIQE rating is more balanced between the classes, providing better distinction between the visual qualities resulting from different network conditions. Second, the PIQE scale more accurately reflects user-perceived visual quality, as it is based on experimental analysis involving human ratings of images with a variety of distortions.
Table~\ref{tab:piqe_confusion_matrix} presents the confusion matrix of PIQE ratings, excluding unobserved classes.
The model's performance within the 'Good' and 'Fair' quality ratings, achieving over 90\% accuracy, indicates the effective utilization of PIQE scores in distinguishing higher-quality images in WhatsApp resulting from varying network conditions. Conversely, the classification of 'Poor' quality demonstrates lower accuracy, with only 64\% of samples predicted correctly, while the majority of the remaining samples are predicted as 'Fair'. Two potential factors are identified for this observation. First, the PIQE CDF in figure~\ref{fig:cdf_piqe} shows a steeper incline in the range of 46 to 53, indicating a higher distribution around the transition value between these two classes. Second, the 'Poor' class contains 10\% of the data, whereas the 'Fair' class contains 36\%. The combination of these factors may introduce a bias towards classifying 'Poor' rated samples as 'Fair'. Across the entire dataset, the UDP model predicted the PIQE score with a mean absolute error (MAE) of 2.1 and classified the PIQE quality rating with an accuracy of 90.2\%.
\begin{table}[t]
\caption{Classification Accuracy of BRISQUE and PIQE Ratings for RTP and UDP Models Under Various Network Conditions}
\centering
\begin{tabular}{|c|c|c|c|c|}
\hline
& \multicolumn{2}{c|}{\textbf{Brisque}} & \multicolumn{2}{c|}{\textbf{PIQE}} \\
\cline{2-5}
& \textbf{RTP} & \textbf{UDP} & \textbf{RTP} & \textbf{UDP} \\
\hline
\textbf{Bandwidth Limits} & 86.1\% & 85.18\% & 88.9\% & 86.74\% \\
\hline
\textbf{Bandwidth Drops} & 93.94\% & 92.82\% & 85.55\% & 85.10\% \\
\hline
\textbf{Packet Loss} & 99.62\% & 99.52\% & 98.82\% & 98.92\% \\
\hline
\end{tabular}
\label{tab:quality_scale_rating_accuracy}
\end{table}

\begin{table}[b]
\caption{Confusion Matrix for PIQE Rating Classification by UDP Model}
\centering
\begin{tabular}{|c|c|c|c|}
\hline
\textbf{Actual} & \multicolumn{3}{|c|}{\textbf{Predicted}} \\
\cline{2-4}
& \textbf{Good} & \textbf{Fair} & \textbf{Poor} \\
\hline
\textbf{Good} & 93.3\% & 6.6\% & 0.1\% \\
\hline
\textbf{Fair} & 4.9\% & 90.0\% & 5.1\% \\
\hline
\textbf{Poor} & 0.2\% & 35.8\% & 64.0\% \\
\hline
\end{tabular}
\label{tab:piqe_confusion_matrix}
\end{table}

\section{Conclusion}
To provide satisfactory services and manage resources efficiently, network operators need to monitor user QoE. The increasing number of video call applications necessitates a general solution, presenting a challenging task. Some applications do not provide QoE ground truth metrics, whereas others require application-specific methods to infer these metrics. In this work, we address this problem by introducing an application-agnostic methodology for estimating QoE metrics from encrypted traffic. Independent of the video application, We infer three objective QoE ground truth metrics that collectively reflect both spatial and temporal video quality. We extend existing solutions and leverage ML models to predict QoE metrics from encrypted traffic. We validate our approach for WhatsApp IMVCA. For that purpose, we generated a dataset of WhatsApp video sessions under various network conditions, including encrypted traffic data and our QoE ground truth metrics. We evaluate the performance of our solution, identify the model's limitations, and investigate WhatsApp's QoE across various network conditions.
Our model achieved high accuracy in predicting QoE metrics across the entire dataset. Specifically, the model predicted the PIQE score with a MAE of 2.1 and classified the PIQE quality rating with 90.2\% accuracy. Additionally, the model predicted the FPS with a MAE of 1.7, achieving an accuracy of 85.2\% within an error tolerance of up to two FPS.
Future work could extend our study by leveraging our metrics to infer more advanced QoE metrics, both subjective and objective. Additionally, improving our frame rate calculation method could simplify data collection from home users, thereby enhancing the overall applicability of our approach. 


\end{document}